\documentstyle[twocolumn,aps,epsfig]{revtex}

\begin{document}

\title{Combination of improved multibondic method
 and the Wang-Landau method}
\author{Chiaki Yamaguchi and Naoki Kawashima}

\address{Department of Physics, Tokyo Metropolitan University,
Hachioji, Tokyo 192-0397, Japan}

\date{\today}

\maketitle

\begin{abstract}
We propose a method for Monte Carlo simulation
  of statistical physical models with discretized energy.
The method is based on several ideas including
  the cluster algorithm,
  the multicanonical Monte Carlo method 
  and its acceleration proposed recently by Wang and Landau.
As in the multibondic ensemble method proposed by Janke and Kappler,
  the present algorithm performs a  random walk in the space 
  of the bond population to yield the state density
  as a function of the bond number.
A test on the Ising model 
  shows that the number of Monte Carlo sweeps
  required of the present method for obtaining the density of state 
  with a given accuracy is proportional to the system size, 
  whereas it is proportional to the system size squared
  for other conventional methods.
In addition, the new method shows a better performance than
the original Wang-Landau method in measurement of physical quantities.
\end{abstract}

%%%%%%%%%%%%%%%%%%%%%%%%%%%%%%%%%%%%%%%%%%%%%%%%%%%%%%%%%%%%%%%%%%%%%%%%%%%%%%%

\section{Introduction}

The Monte Carlo simulation is one of the most powerful tools for 
  investigating models in statistical physics\cite{Mon}.
Although the Metropolis method\cite{M1} and its variations are
  available for simulating variety of models,
  they are not necessarily the best methods
  when the system of interest has a strong long-ranged correlation.
Essentially two approaches have been proposed for overcoming the
  drawbacks of such local-updating methods.
In one approach, one uses an ensemble entirely different from 
  the ordinary canonical ensemble with a fixed temperature,
  whereas in the other approach one extends the original ensemble by
  introducing auxiliary variables.

Multicanonical method\cite{Berg1,Berg2,Lee1}, broad histogram
  method\cite{O1}, and the flat histogram method\cite{W1}
  belong to the first category.
In these methods a random walk in the energy space is performed
  to calculate the state density as a function of the energy.   
Multicanonical method was applied to the $Q$-state Potts model, for example,
  and turned out very successful\cite{Berg1}.
Meanwhile it was realized that the random walker tends to be
  blocked by the edge of the already visited area.
In addition, because of the general feature of random walks,
  it takes a long time to go from one end of the area to the other.
Recently Wang and Landau\cite{WL} succeeded in removing
  these problems by penalizing moving to and staying at the energy
  which has been visited many times.
The efficiency of the Wang-Landau (WL) method was also demonstrated
 in an application to antiferromagnetic $Q$-state Potts model on a 
 simple cubic lattice\cite{YamaguchiOkabe2001}.
In particular, the method turned out to be powerful in studying 
 the ground state properties due to the fast diffusion accelerated
 by the WL method.

The second category includes various cluster algorithms.
In cluster algorithms, graph degrees of freedom are introduced to
 extend the original ensemble.
In most of their successful applications,
  clusters of the size of correlation length\cite{SW1,Wolff1} are 
  formed and flipped.
A cluster algorithm is applied to $Q$-state Potts model and 
  proven much more efficient\cite{SW1} than local updating algorithms. 

Janke and Kappler\cite{J1} proposed the multibondic method, the
  combination of the multicanonical method and the cluster algorithm. 
They took the two-dimensional $Q$-state Potts model as an example.
They measured the computational time required for the random walker to traverse
  the interval between the two peaks in the canonical probability distribution.
The traverse time in units of Monte Carlo sweeps
  was found to be proportional to the number of spins, $N^1$, 
  whereas it is proportional to $N^{1.3}$ for the ordinary 
  multicanonical method,
  where $N$ is the total number of spins in the system.
The multicanonical cluster algorithm\cite{R} shows the same 
  size dependence as the Janke-Kappler algorithm (referred to as
 `JK' in the present paper). 
  The comparison between the Janke-Kappler algorithm 
  and this algorithm can be found in \cite{J1}.  

In this paper, we propose a new method based on the JK
method (or, more generally, the multibondic ensemble method) and
the Wang-Landau acceleration method.
In Section II, we briefly review the multibondic ensemble method.
In Section III, we propose a modification of the Janke-Kappler method (MJK)
to avoid the possible lasting effect of the initial graph.
Then, in Section IV, a combination of the MJK with 
the Wang-Landau method (WL) is discussed.
We refer to this combined method as MJKWL.
In Section V, we demonstrate the efficiency of the MJKWL 
by comparing it with other methods.
In particular, it is shown that the MJKWL is better than
each of its ingredients, i.e., the JK method and the WL method.
In Appendix A, we present a simple and exact relationship 
between the density of state (DOS) as a function of the energy 
and the DOS as a function of the bond number for the $Q$-state Potts
model in any dimensions. 

Since we are forced to use many abbreviations in the present paper
to refer to various methods, 
it may be convenient to summarize all of them here:
\begin{description}
\item[JK] ... The Janke-Kappler method.
\item[WL] ... The Wang-Landau method.
\item[SWL] ... The Wang-Landau method with single spin update, i.e.,
               the original Wang-Landau method.
\item[MJK] ... The modified Janke-Kappler method.
\item[MJKWL] ... The modified Janke-Kappler method with the Wang-Landau
                 acceleration method.
\end{description}

%%%%%%%%%%%%%%%%%%%%%%%%%%%%%%%%%%%%%%%%%%%%%%%%%%%%%%%%%%%%%%%%%%%%%%%%%%%%%%%

\section{The multibondic ensemble method}
\label{MultiBondic}
Since our method can be viewed as a derivative of  
  the Janke-Kappler algorithm, 
  we give a brief review of the multibondic ensemble method.
In the following, we take the $Q$-state Potts model as an example 
  to make our description concrete.
However, the generalization to other models with discrete 
  energy is straightforward, and we try to describe our method
  so that the generalization appears obvious.

The Hamiltonian is given by
$$
  {\cal H} = - J \sum_{\langle i j \rangle} \delta_{\sigma_i, \sigma_j},
  \quad \sigma_i = \{ 1, \cdots , Q\} 
$$
where $J$ is the exchange coupling constants and $\langle i j \rangle$
denotes a nearest neighbor pair.
In what follows, we take $J$ as the unit of the energy,
and $J/K_B$ as the unit of temperature,
 where $k_B$ is Boltzmann's constant.
We first represent the partition function as a double
  summation over states $S$ and graphs $G$,
  following the general framework of 
  the dual algorithm\cite{KD,KG};
\begin{equation}
Z (T) = \sum_S W_0 (S) = \sum_{S,G} W_0(S,G) 
\equiv \sum_{S, G} V_0 (G) \, \Delta (S, G)
\end{equation}
This is nothing but the well-known Fortuin-Kasteleyn 
  representation\cite{FK1}.
$W_0(S)$ is the weight of state $S$, whereas $\Delta (S, G)$ is 
  a function that takes the value one when $S$ is compatible to $G$
  and takes the value zero otherwise. 
$V_0(G)$ denotes the weight for graph $G$ defined as
$$
  V_0(G) = V_0(n_b(G),T) \equiv (e^{1/T} - 1)^{n_b(G)}
$$
  where $n_b(G)$  is the number of bonds in $G$,
  in the case of the $Q$-state Potts model.
Although the only graph elements are bonds in this case,
  a graph consists of more than one type of elements in general applications.
Therefore, in more general terms, $n_b(G)$ is a $p$-dimensional 
  vector variable $i$-th element is the number of graph elements 
  of the $i$-th kind contained in the graph $G$.
By taking the summation over $S$ and $G$, fixing the fixed number of the bond, 
 the above expression for the partition function is reduced to
\begin{equation}
Z (T) = \sum_{n_b = 0}^{N_P} \Omega (n_b) V_0 (n_b, T) . \label{eq:d_func_nb}
\end{equation}
where $N_P$ is the total number of nearest neighbor pairs in the whole
 system ($N_P = d N = d L^d$ for $d$-dimensional hyper cubic lattices). 
 Here, $\Omega (n_b )$ is the DOS of the bond number defined as
  the number of consistent combinations of
  graphs and states such that the graph consists of $n_b$ bonds;
$$
\Omega (n_b) \equiv \sum_{\{G|n_b (G) = n_b\}} \sum_S \, \Delta (S, G).
$$

In the multibondic ensemble method, we replace $V_0(n_b,T)$ by a 
free function, which we denote by $V(n_b)$.
By adjusting this function, we try to make the histogram flat
as a function of $n_b$.
In other words, $V(n_b)$ is adjusted so that the product
of $\Omega(n_b)$ and $V(n_b)$ may be independent of $n_b$.
Since initially $\Omega(n_b)$ is unknown, we achieve this adaptively.
The way how we modify the trial weight is
described in the following sections.

There are several different ways for adjusting the trial weight.
The original JK method is one of them.
The detailed description of the JK method can be found in Appendix B
and in the original paper\cite{J1}.

%%%%%%%%%%%%%%%%%%%%%%%%%%%%%%%%%%%%%%%%%%%%%%%%%%%%%%%%%%%%%%%%%%%%%%%%%%%%%%%

\section{Modification of the Janke-Kappler algorithm}

As stated in Appendix B, in the original JK algorithm 
the initial graph for each Monte Carlo sweep may have a long 
lasting effect upon the 
subsequent states during the sweep since the bond update
is done only sequentially.
In this section, we propose a modification of the Janke-Kappler algorithm
(MJK) to reduce the possible lasting effect of the initial graph
as much as possible.

We first choose any consistent combination of a state and a graph 
as an initial condition.
For the initial choice of $V(n_b)$, we choose $V (n_b) = 1$ for any $n_b$.

In each Monte Carlo sweep of the MJK, we choose the number of the bonds 
to be placed on the whole system before actually placing them.
The number $n_b$ is chosen with the following probability,
$$
P(n_b|n_p(S)) \propto 
\left( \begin{array}{c} n_p(S) \\ n_b \end{array} \right) V(n_b).
$$
Here, $n_p(S)$ is the number of ``satisfied'' pairs in the current state $S$, 
i.e.,
$$
n_p(S) = \sum_{\langle i j \rangle} \delta_{\sigma_i(S), \sigma_j(S)}. 
$$
and
$$
\left( \begin{array}{c} l \\ m \end{array} \right) \equiv
 \frac{l !}{m ! (l - m) !} . 
$$
Note that the choice of $n_b$ is based only on the information of 
the current state $S$ at the beginning of the sweep.
As for the graph $G$ at the beginning of the sweep, 
we simply delete all the bonds in it.
We then choose $n_b$ pairs at random out of $n_p (S)$ satisfied ones
and place new bonds on them.
It is clear that there is no direct influence of the initial graph
on the final graph. The correlation between them arises only through the state which should
be compatible to the initial graph.
After the placement of the $n_b$ bonds, we `flip' all the clusters of sites
 with probability one half
where flipping a cluster means changing all the variables on it simultaneously.
This completes one Monte Carlo `sweep'.
It is easy to show that this procedure satisfies the extended detailed
balance condition\cite{KG}: 
$$
  P(G|S)W(S) = P(S|G)W(G)
$$
where $W(G) \equiv \sum_S W(S,G)$ and $W(S,G) \equiv V(G)\Delta(S,G)$.

Another modification should be done in order to make MJK method
better than the original JK method.
In the JK method,
the histogram $H(n_b)$ is updated by the simple rule
\begin{equation}
  H(n_b) \Leftarrow H(n_b) + \delta(n_b,n_b(G))
  \label{UpdatingRule}
\end{equation}
every time a part of graph, i.e., a bond on a pair of sites, is updated.
It means that we update $H(n_b)$ in a sweep as many times as 
the number of bonds.
Although the successive values of $n_b$ in the same sweep
are strongly correlated with each other in the JK algorithm,
one can still get statistically more informative data
by taking them all into account.
It is roughly equivalent to adding some smooth function to $H(n_b)$ 
at every Monte Carlo sweep, in contrast to adding a delta function.

However, in the MJK method, $H(n_b)$ is updated only
once in every Monte Carlo sweep, which means that a delta function
is added to $H(n_b)$ at each sweep according to the
the updating rule Eq. (\ref{UpdatingRule}).
In order to remove this disadvantage, in the MJKWL method,
we add to the histogram the expectation values of the delta 
function $\delta(n_b,n_g(S))$, rather than the delta function itself.
The resulting updating rule for $H(n_b)$ is
$$
  H(n_b) \Leftarrow H(n_b) + N_{\rm P}\, P(n_b|n_p(S)).
$$
Although including a constant $N_P$ is not relevant,
it is added in order to make the magnitude of the histogram 
comparable to the one in the JK method.

%%%%%%%%%%%%%%%%%%%%%%%%%%%%%%%%%%%%%%%%%%%%%%%%%%%%%%%%%%%%%%%%%%%%%%%%%%%%%%%

\section{Combining with the Wang-Landau method}
\label{WangLandau}

It has been known since the first proposal of the multicanonical method
that the random walker tends to be stuck at the boundary
which separates the region visited already from the one not visited yet.
This difficulty has been removed by the recent technique proposed by
Wang and Landau\cite{WL}.
Their method seems to be useful also in accelerating the diffusion
of the random walker.

In the WL method, the DOS phase (see the next section)
of the computation consists of varying number of consecutive 
sets of simulation, as is also the case with the ordinary multi-canonical
method and its derivatives.
However, the important difference lies in the way the trial weight is
updated.
In the conventional multi-canonical methods, the trial weight
is updated only at the end of each set of simulations.
During each set, the trial weight and, consequently, the transition 
probability are fixed.
In the WL method, on the other hand, every time the state of the system
is renewed, the trial weight $W(E)$ is updated as
$$
  \ln W(E) \Leftarrow \ln W(E) - \lambda \delta(E,E(S)) \qquad (\lambda >0),
$$
where $E(S)$ is the energy of the current state.
The positive parameter $\lambda$ is introduced to control the magnitude
of the expelling force imposed on the random walker.
If the parameter is large, the random walker quickly moves out of
the region which it has already visited.
In other words, the histogram is forced to be flat by
 this parameter.  
However the very presence
of this force breaks the detailed balance and therefore makes 
the resulting trial weight differ from the
correct one, i.e., the inverse of the DOS.
On the other hand, if it is small, the resulting trial weight
is reliable while the convergence tends to be slow.
Therefore, the general strategy is to set this value large initially
and make it smaller as the trial weight approaches the correct one.

The value $\lambda$ is kept constant throughout each set of simulation
and is reduced by some factor at the beginning of the next set of simulation.
Wang and Landau suggested $1 / 2$ for the reduction factor.
Each set of simulation terminates when the histogram of the set
satisfies some predetermined condition concerning its flatness.
The histogram is reset at the beginning of a new set while
the trial weight $W(E)$ is not.
The whole calculation is terminated when $\lambda$
becomes smaller than a predetermined value.

In order to combine the WL method with the MJK method described in
the previous section, we may simply replace the energy space 
in the original WL method by the bond-number space.
To be more specific, 
$H(E)$ is replaced by $H(n_b)$,
$W(E)$ by $V(n_b)$, and
$\delta(E,E(S))$ by $\delta(n_b,n_b(G))$.
However, as for the updating rule of $V(n_b)$,
$$
  \ln V(n_b) \Leftarrow \ln V(n_b) - \lambda \, N_P \, P(n_b|n_p(S))
$$
is the better choice than
$$
  \ln V(n_b) \Leftarrow \ln V(n_b) - \lambda \delta(n_b , n_b(G))
$$
for the same reason as we stated in the previous section.
We therefore use the former updating rule in the MJKWL method
for the sample calculation presented below.

\newpage

%%%%%%%%%%%%%%%%%%%%%%%%%%%%%%%%%%%%%%%%%%%%%%%%%%%%%%%%%%%%%%%%%%%%%%%%%%%%%%%

\section{Efficiency of the method}

We now discuss the performance of the above-mentioned methods.
Since the present method (MJKWL) is the combination of MJK and WL, 
it should be demonstrated that this combination is meaningful, i.e.,
MJKWL is qualitatively better than both of the two ingredients.

First it should be noted that there are several measures of
performance.
In the WL method, 
such as MJKWL and SWL, 
the whole computation process consists of two phases; 
the DOS phase and the measurement phase.
In the DOS phase, the computation is performed mainly
for obtaining an estimate of the DOS.
During this phase, several sets of simulation are done for
the adoptive adjustment of the fictitious weight
(the trial weight $V(n_B)$ in Section \ref{MultiBondic} for
the multibondic methods such as JK, MJK and MJKWL
and the trial weight $W(E)$ in Section \ref{WangLandau} for 
 the Wang-Landau method with single spin update (SWL)).
At the end of this phase, some of the physical quantities,
such as the entropy, the energy and the specific heat, can be 
computed with the resulting DOS.
For other quantities, however, some additional simulation
should be performed with a fixed trial weight and with the
controlling parameter $\lambda$ set to be zero.
We call this part the measurement phase.
(In the case where the value of $\lambda$ for the last set
of simulation in the DOS phase is negligibly small,
a separate measurement phase may not be necessary. 
In such cases, we regard the last set as the measurement phase.)
In what follows, we discuss the computational time required 
for the DOS phase and that for the measurement phase, separately.
As we see below, the DOS is obtained much faster in the DOS phase
 of MJKWL than in JK and MJK,
 while SWL shows qualitatively the same performance as MJKWL.
The difference between MJKWL and SWL can be seen in the
measurement phase.
Namely, MJKWL yields much better statistics in the measurement 
phase than SWL within the same number of Monte Carlo sweeps.

A remark should be placed here concerning the sources of errors
in the two phases.
During the DOS phase, the systematic error as well as the statistical
error is present. 
The systematic error is due to the obvious fact that there may be a
region which the random walker has not visited yet.
In the methods based on the  Wang-Landau acceleration,
the fact that the controlling parameter $\lambda$ is not zero
 in another source of systematic error.
Because of this systematic error, the dependence of the
total error on the duration of the simulation is complicated.
On the other hand, in the measurement phase 
there is no other sources of errors than
the ordinary statistical ones.
Therefore the precision of the result in this phase
is proportional to the inverse of the square root of the number
of Monte Carlo sweeps.

In what follows, we argue and demonstrate that the methods 
without the Wang-Landau acceleration, such as JK and MJK,
require the number of Monte Carlo sweeps of the order $O(N^2)$ whereas
the methods with the Wang-Landau acceleration, such as MJKWL and SWL
require $O(N)$ to achieve the same accuracy in the DOS.

In all the methods, discussed here,
we start with some ad-hoc initial guess for the DOS.
Then, the resulting histogram has a rather narrow range of distribution.
Therefore, in order to make the histogram flat throughout the whole
energy (or bond number) range, we have to repeat simulations.
Every time we start a new set of simulation, we improve the
initial guess for the density of states based upon the outcome
of the last set of simulation.
The difficulty arises near the boundary between the two regions; 
the region which has been visited already in the previous simulations
and the region which has not.
When the random walker in the energy (or the bond number) space
hits the boundary during the simulation,
it usually bounces back and, even if it does not,
it seldom goes far beyond the boundary.
Therefore, the width of the visited region increases
by only a few steps as a result of the whole set.
It follows that the number of sets of simulation required for
making the histogram flat is proportional to the width of the
energy (or bond number) space, that is, $O(N)$.
In addition, each set must be long enough for the walker to traverse
the whole previously visited region.
Since the width of the previously visited region is of the 
order $O(N)$ in general and the typical distance the
walker traverse in a single Monte Carlo sweep is $O(N^{1/2})$,
the number of Monte Carlo sweeps required for the walker to traverse 
the region is $O((N/N^{1/2})^2)=O(N)$.
These factors are multiplied to make the total number of sweeps
required for the whole DOS phase of the order $O(N^2)$.

In contrast, in methods with Wang and Landau's acceleration, 
the situation described above cannot happen.
This is because the current histogram affects the current 
weights and transition probabilities,
 such that the weights for the frequently visited positions become smaller.
This forces the walker to move out of the already visited region
 and make histogram flat.
 Since this situation is similar to the one-dimensional self-avoiding walk,
 a natural guess is that the number of steps (i.e., the number of local
 updatings) required for the walker to traverse the already visited region 
 is proportional to the size of the region, which is $O(N)$.
 In units of sweeps, it is $O(N^0)$.

To check if this simple argument is correct, we performed simulations
for ferromagnetic Ising model on a square lattice
using three different methods: MJKWL, MJK and JK.
For these three methods, 
we set an initial weight $V (n_b ) = 1$ for all $n_b$. 
We measured the number of Monte Carlo sweeps required 
for obtaining the DOS with a roughly fixed precision,
as a function of system size $N\equiv L^2$.
It should be remarked here that we cannot rigidly fix 
the target precision of the DOS because the termination
condition in MJKWL is defined in terms of the flatness of the
histogram and the value of the controlling parameter $\lambda$,
not the number of Monte Carlo sweeps nor the precision of the DOS.
Therefore, we performed a MJKWL simulation first with
some reasonable choice of the termination condition.
Then, we performed simulations using JK and MJK.
The current estimate of the DOS is updated frequently in these
simulations so that the simulation can be terminated as soon as
the precision of the DOS estimate reaches the same as that 
obtained in the MJKWL simulation.
The precision of the DOS is measured by the following quantity.
$$
  \epsilon (L) \equiv \frac1{N_P + 1} \sum_{n_b=0}^{N_P} 
  \left|
  \ln \Omega(n_b) - \ln \Omega^{{\rm (exact)}}(n_b)
  \right|
$$
The exact DOS, $\Omega^{{\rm (exact)}}$,
is obtained through Eq. (\ref{eq:Og}) and
the exact DOS as a function of the energy\cite{B1}. 
In what follows, the termination condition for the MJKWL
is the same for all the system sizes. 
It turned out that the resulting signal-noise ratio of the DOS, 
$\epsilon (L)$, is roughly independent of the system size.

Our procedure for the MJKWL simulation is as follows.  
The reduction factor $\lambda$ is divided by two when
 each set of simulation is terminated.
Each set is terminated when
 the smallest $H (n_b)$ becomes greater than $0.8$ times 
 the average value of $H (n_b)$. 
The whole calculation is terminated when $\lambda$ becomes 
less than $10^{- 8}$.
This procedure is essentially the same as suggested 
in the original paper by Wang and Landau\cite{W1}
except that we work with the DOS as a function of $n_b$ rather than $E$.

For MJK and JK, we perform a number of subsequent sets of simulations 
 to improve the estimates of the DOS.
We start with a relatively short set and gradually
 make it longer.
The way we increase the number of sweeps of a set
 depends upon whether the random walker has already visited the whole
 bond-number space. 
If $n_b$ has not visited the whole $n_b$ space at the end of the
$i$-th set, the number of sweeps for the $i+1$-th set is chosen as
$$
  t_{i+1} = 10 (m_i + \sqrt{N_P})
$$
where $m_i$ is the number of the already visited values of $n_b$
in the $i$-th set. 
If the above argument is correct, i.e., the random walker moves
in the bond-number space as a self-avoiding walker,
this choice of $t_{i+1}$ should give the walker an enough time
to traverse the whole region of previously visited values of $n_b$
and touch the boundary a few times.
Therefore it should be enough to expand the visited region.
If $n_b$ has already visited the whole $n_b$ space at the end of the
$i$-th set, the number of sweeps for the $i+1$-th set is given by
$$
  t_{i+1} = 2 \, t_i.
$$
This choice will provide the walker with an enough time to develop 
appreciably better trial weights than the previous sets.
The whole procedure yields the total number of sweeps of
the order $O(N^2)$ if the above argument is correct. 
The entire process is terminated when the estimated 
$\Omega (n_b)$ has become as good as that obtained with the MJKWL.

The computation is done for system sizes $L=4,8,16,24$ and $32$
 as all other sample calculations presented below.
The results is shown in Figure $1$.
The average is taken over about 30 independent simulations.
We can easily see that MJKWL is the best method
among the three multi-bondic methods for larger $N$.
It can be also seen that the MJK is better than the JK. 
Two lines are drawn in Figure $1$ for references.
The lower dashed line corresponds to $t \propto O(N^1 )$ whereas the upper
dashed line $t \propto O(N^2 )$
We can see that the MJKWL requires $O (N^1 )$ sweeps while 
the MJK and the JK require $O (N^2)$ sweeps, as expected from the
argument.
We have also confirmed that the relative statistical error in the DOS
obtained by MJKWL does not strongly depend on the system size. 

\begin{center}
\begin{figure}
\epsfxsize=0.65\linewidth
\centerline{\epsfbox{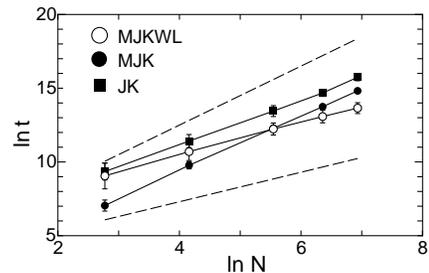}}
\caption{
 The total number of Monte Carlo sweeps performed to obtain the
 same accuracy in the DOS estimate, as a function of the system
 size $N\equiv L^2$
 for ferromagnetic Ising model on a square lattice.
 Three methods are examined:
 the modified Janke-Kappler algorithm with the Wang-Landau method (MJKWL),
 the modified Janke-Kappler algorithm (MJK),
 and the ordinary Janke-Kappler algorithm (JK).
 The average is taken over $30$ independent runs.
 The upper and lower lines are for references,
 corresponding to $t \propto O (N^2 )$ and $t \propto O(N^1 )$,
 respectively.
}
\label{Comparison}
\end{figure}
\end{center}

The performance of SWL, i.e., Wang and Landau's original method
using single spin-flips, is also examined.
We set the initial weight $W (E) = 1$ for all $E$.
We measured the total number of Monte Carlo sweeps as a function
of the system size.
Again the precision of the resulting estimate of the DOS 
does not strongly depend on the system size.
The result is shown in Figure $2$.
We can see that the SWL require $O (N^1 )$ sweeps for large systems.
Therefore, it can be concluded that the SWL has the same qualitative
performance as the MJKWL in the DOS phase.

\newpage

\begin{center}
\begin{figure}
\epsfxsize=0.65\linewidth
\centerline{\epsfbox{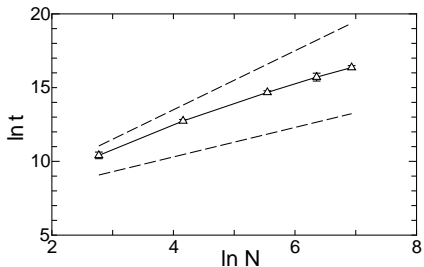}}
\caption{
 The total number of Monte Carlo sweeps
 as a function of the system size $N\equiv L^2$
 for ferromagnetic Ising model on a square lattice.
 The calculation is performed following the Wang and Landau's
 original procedure.
 The average is taken over $10$ independent runs.
 The upper and lower lines are for references,
 corresponding to $t \propto O (N^2 )$ and $t \propto O(N^1 )$,
 respectively.
}
\label{SWL}
\end{figure}
\end{center}

To compare the efficiency of MJKWL and SWL in the measurement phase,
 we calculate the magnetization squared, $M^2$
 divided by $N^2$ for ferromagnetic Ising model on a square lattice. 
 We first estimate the DOS in the DOS phase.
 Using this DOS, 
 we then perform 50 independent runs for the measurement phase
 using different random number sequence for each run.
 Each run consists of $100 \times N$ sweeps, and produces a
 histogram and a set of microcanonical averages of the squared
 magnetization, as is usually done in any multicanonical-type
 methods.
 Based on these information, the canonical average of the squared 
 magnetization at the critical temperature is computed for each run.
 Then, we compute the standard deviation of these 50 canonical
 averages.
 This standard deviation is proportional to the statistical error
 in the final estimate
 and can be used as a measure of the efficiency with which
 the spin configuration is updated during the simulation.

The result is shown in Figure $3$.
 As is clear from the figure, MJKWL is better than SWL.
 The difference in the standard deviation tends to increase as the
 system becomes larger.
 This is because the spin configuration is updated by clusters in MJKWL
 whereas it is updated by single spins in SWL.
 Therefore, the configuration is decorrelated much faster in MJKWL
 than in SWL.
 To be more specific, a random walker in SWL must visit states
 with very different values of energy in order to visit a state with 
 very different value of the magnetization,
 whereas a random walker in MJKWL does not have to because
 the state can change even without changing the bond number at all.

\begin{center}
\begin{figure}
\epsfxsize=0.65\linewidth
\centerline{\epsfbox{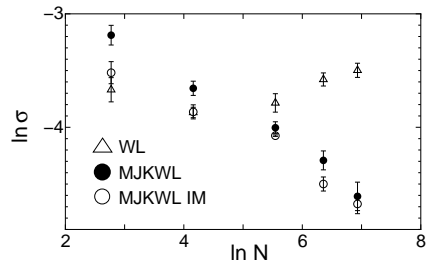}}
\caption{
 The standard deviation of $50$ independent estimates of the
 squared magnetization per spin thermally averaged at
 the critical temperature, $T_C = 2 / \ln (1 + \sqrt{2})$,
 for the ferromagnetic Ising model on a square lattice.
For each run, $100 \times N$ sweeps are performed.
}
\label{Sigma}
\end{figure}
\end{center}

%%%%%%%%%%%%%%%%%%%%%%%%%%%%%%%%%%%%%%%%%%%%%%%%%%%%%%%%%%%%%%%%%%%%%%%%%%%%%%%

\section{Summary}

We have proposed a combination of the Janke-Kappler algorithm with
the Wang-Landau acceleration method, together with 
a modification of the Janke-Kappler algorithm.
The number of Monte Carlo sweeps
required for obtaining the DOS with several methods 
have been measured and compared.
It has been demonstrated
that the number of Monte Carlo sweeps required
 for obtaining the DOS in the methods without the Wang-Landau acceleration
is proportional to $N^2$, 
whereas in the present method (MJKWL)
it is proportional to $N^1$. 
The new method is also compared with Wang and Landau's
original method based on single spin updating (SWL).
The result shows that the spin configuration is much more
efficiently updated in MJKWL than in the SWL.

The proposed modification to the Janke-Kappler algorithm 
 turns out to be useful in reducing the
CPU time requirement, though not as vital as the Wang and 
Landau's idea in the cases shown in the present article.

We have also deduced an exact relation between the DOS
as a function of the energy and that as a function of
the bond number for the $Q$-state Potts model in any dimensions.

The present method can be easily extended to other models with
discrete degrees of freedom, in particular, when the cluster
algorithm has been already devised.
Quantum spin models can also be dealt with in the present scheme.
In a loop-cluster algorithm\cite{LoopCluster},
the partition function is expressed as a sum of classical (non-quantum)
weight over spin configurations and graphs.
The graph degrees of freedom can be divided into a continuous part
(the locations of the graph elements in the imaginary time axis)
and a discrete part (the number and the types of the graph elements).
The present scheme can be applicable to the latter discrete part
of graph degrees of freedom.
The work in this direction is now under progress and will be reported
elsewhere\cite{FutureWork}.

%%%%%%%%%%%%%%%%%%%%%%%%%%%%%%%%%%%%%%%%%%%%%%%%%%%%%%%%%%%%%%%%%%%%%%%%%%%%%%%

\section{Acknowledgment}
We thank J.-S.\ Wang, Y.\ Okabe, H.\ Otsuka and Y.\ Tomita
 for useful comments.

%%%%%%%%%%%%%%%%%%%%%%%%%%%%%%%%%%%%%%%%%%%%%%%%%%%%%%%%%%%%%%%%%%%%%%%%%%%%%%%

\section*{Appendix A: Relationship between two densities of states}

We derive the exact relationship between $g (E)$ and $\Omega (n_b)$.
We first define a parameter $x$ as;

$$
x \equiv \exp (1 / T).
$$
Using this $x$, we can express the partition function as 
$$
Z (x) = \sum^0_{E = - N_P} g (E) \, x^{- E}.
$$
In terms of the number of bonds, it is written as
$$
Z (x) = \sum^{N_P}_{n_b = 0} \Omega (n_b) \, (x - 1)^{n_b}.
$$
The range of the energy is $- N_P \le E \le 0$ and
that of the number of bonds is $0 \le n_b \le N_P$
where $N_P$ is the total number of nearest neighbor pairs of spins.
Differentiating the above two equations $l$
times with respect to $x$, we obtain
\begin{eqnarray}
\frac{\partial^l}{\partial^l \, x}
 Z (x) &=& l! \, g (- l) + \frac{(l + 1)!}{1!} \, g (- l - 1) \, x 
 \nonumber \\
 &+& \frac{(l + 2)!}{2!} \, g (- l - 2) \, x^2 + \cdots, 
\nonumber \\
\frac{\partial^l}{\partial^l \, x}
 Z (x) &=& l! \, \Omega (l) + 
\frac{(l + 1)!}{1!} \, \Omega (l + 1) \, (x - 1) \nonumber \\
 &+& \frac{(l + 2)!}{2!} \, \Omega (l + 2) \, (x - 1)^2 + \cdots. 
\label{TwoPartitionFunctions}
\end{eqnarray}
By comparing these two equations after taking
the limit $T \to \infty$, ($x \to 1$), we arrive at 
the relation of $\Omega (l)$ and $g (l)$: 
$$
\Omega (l) =  g (- l) + \frac{(l + 1)!}{l! 1!} \, g (- l - 1) 
 + \frac{(l + 2)!}{l! 2!} \, g (- l - 2) + \cdots, \label{eq:l2}
$$
or
\begin{equation}
\Omega (n_b) = \sum_{j=0}^{N_P - n_b} 
\left( \begin{array}{c} n_b + j \\ j \end{array} \right)
g (- n_b - j).  \quad (0 \le n_b \le N_P) 
\label{eq:Og}
\end{equation}
By setting $x=0$ in Eq. (\ref{TwoPartitionFunctions}), we obtain
\begin{eqnarray}
g (E) &=& \sum_{j=0}^{d \, N + E} (-1)^j 
\left( \begin{array}{c} - E + j \\ j \end{array} \right)
\Omega (- E + j), \nonumber \\ &\qquad& \qquad (-N_P \le E \le 0).
\label{eq:gO}
\end{eqnarray}
Equation (\ref{eq:Og}) is useful for
obtaining $\Omega (n_b )$ from $g (E)$ such as 
those obtained by Beale\cite{B1}. 
However, computing $g (E)$ from $\Omega (n_b)$
using equation (\ref{eq:gO}) is not practical when the estimates of 
$\Omega(n_b)$ contains statistical error,
because the $(-1)^j$ factor in equation (\ref{eq:gO})
magnifies the relative magnitude of the errors.

Using Eq. (\ref{eq:gO}), we can obtain, for example, the expression for the
ground state entropy,
$$
  e^{S_0} \equiv g(-N_P) = \Omega(N_P).
$$
It should be remarked that 
the direct outcome of the actual simulation is not $\Omega(n_b)$
itself but the relative magnitude of $\Omega(n_b)$'s.
Therefore, in order to obtain an estimate of $\Omega(N_P)$,
we have to use the fact that
$$
  \Omega(0) = Q^N
$$
for the $Q$ state Potts model.
With this equation, the absolute magnitude of $\Omega(n_b)$ 
can be determined.
In other words, if $\tilde\Omega(N_P)$ is the direct outcome
of the simulation and therefore proportional to $\Omega(N_P)$,
the entropy is given by
$$
  e^{S_0} = \frac{\tilde\Omega (N_P)}{\tilde \Omega (0)} \times Q^N.
$$

%%%%%%%%%%%%%%%%%%%%%%%%%%%%%%%%%%%%%%%%%%%%%%%%%%%%%%%%%%%%%%%%%%%%%%%%%%%%%%%

\section*{Appendix B: the Janke-Kappler algorithm}

Here, our implementation of the Janke-Kappler algorithm\cite{J1} is 
described. 
For a given spin configuration and a graph,
we start with making a random choice of a nearest neighbor pair of sites.
With some probability, we remove the bond if there is one 
already on the chosen pair, whereas we place a new bond 
if there is no bond on the pair and if the pair is satisfied, 
 again probabilistically. 
We say a pair $(i, j)$ is satisfied if $\sigma_i = \sigma_j$.
In either case, the probability for updating is
of the heat-bath type;
$$
 P(G'|S,G) \equiv \frac{W(S,G')}{W(S,G)+W(S,G')} 
$$
where $G'$ is the graph in the proposed final state.
$W(S,G)$ is defined as
$$
  W(S,G) \equiv V(G) \Delta(S,G).
$$
where $V(G)$ is the trial weight which is adoptively adjusted.
To be more specific, if there is a bond already on the chosen pair,
we remove it with probability
$$
  \frac{V(n_b-1)}{V(n_b) + V(n_b-1)}.
$$
If there is no bond and if the pair is satisfied, 
 we place a new bond to the pair with probability
$$
  \frac{V(n_b+1)}{V(n_b) + V(n_b+1)}.
$$
If the pair is not satisfied, we leave it unconnected.
We repeat this procedure many times so that every nearest neighbor
pair is chosen and examined once on the average.
After these repetitions, we `flip' all the clusters of sites
with probability one half.

If $V(n_b)$ is simply written as $v^{n_b}$ with some constant $v$,
as is the case with the original weight $V_0$,
the decisions of placing bonds can be made 
for each nearest neighbor pair independently.
In such a case, the resulting algorithm is nothing but the 
Swendsen-Wang algorithm\cite{SW1}.
However, since the adoptively chosen $V(n_b)$ is not in general
factorized, the decisions are dependent.
Therefore, in the original Janke and Kappler's method
only one nearest neighbor pair is examined at a time.
For this reason the graph in the Janke-Kappler algorithm
can change only gradually.
In general, this is a disadvantage because there may be
some unfavorable region or a `barrier' in the bond-number space
which hinders the random walker from moving from one side of it
to the other.
This disadvantage can be removed by the modification proposed in
the main text, in which the random walker can jump from one
side to the other in one step without hitting the barrier.

In a practical applications, we perform some number of
sweeps to obtain a histogram of $n_b$, $H(n_b)$.
Then, we adjust $V(n_b)$ by
$$
  V(n_b) \Leftarrow V(n_b)/H(n_b).
$$
With this new weight, we redo the simulation.
The whole procedure is repeated until $H(n_b)$ becomes
sufficiently $n_b$ independent.

%%%%%%%%%%%%%%%%%%%%%%%%%%%%%%%%%%%%%%%%%%%%%%%%%%%%%%%%%%%%%%%%%%%%%%%%%%%%%%%

%%%%%%%%%%%%%%%%%%%%%%%%%%%%%%%%%%%%%%%%%%%%%%%%%%%%%%%%%%%%%%%%%%%%%%%%%%%%%%%

\end{document}